\documentclass{IEEEtran4PSCC}
\ifCLASSINFOpdf
   \usepackage[pdftex]{graphicx}
\else
   \usepackage[dvips]{graphicx}
\fi
\usepackage[cmex10]{amsmath}

\usepackage{cite}

\usepackage{subcaption}
\usepackage{mleftright}
\usepackage{hyperref}

\usepackage{amsmath}
\usepackage{amssymb}
\usepackage{amsfonts}

 \usepackage{tabularray}

\usepackage{array}
\usepackage{enumitem}
\usepackage{booktabs}
\usepackage{graphicx}
\usepackage{tabularx}
\usepackage{etoolbox}

\usepackage{mdframed}

\newmdenv[leftline=false,rightline=false,linewidth=1pt]{topbot2}

\usepackage{forest}
\usetikzlibrary{external}

\let\oldtikzexternalgetnextfilename\tikzexternalgetnextfilename \renewcommand{\tikzexternalgetnextfilename}[1]{\oldtikzexternalgetnextfilename{#1}\expandafter\tikzsetnextfilename\expandafter{#1}}

\usepackage{pgfplotstable}
\usepackage[outline]{contour}
\contourlength{1.2pt}

\pgfplotsset{compat=1.13} 

\usetikzlibrary{spy}
\usetikzlibrary{calc}
\usetikzlibrary{fadings}
\usetikzlibrary{patterns}
\usetikzlibrary{shadows}
\usetikzlibrary{mindmap}
\usetikzlibrary{backgrounds}
\usetikzlibrary{shapes.symbols}
\usetikzlibrary{shapes.multipart}
\usetikzlibrary{shapes.geometric}
\usetikzlibrary{automata,positioning}
\usetikzlibrary{decorations.fractals} 
\usetikzlibrary{decorations.markings}
\usetikzlibrary{decorations.pathreplacing}
\usetikzlibrary{decorations.pathmorphing}
 
\usepackage{bm}

\usepackage{nomencl}
\makenomenclature
\tikzset{edge from parent/.style={segment angle=10,draw}}

\tikzset{
 my rounded corners/.append style={rounded corners=2pt},
}

\def\BibTeX{{\rm B\kern-.05em{\sc i\kern-.025em b}\kern-.08em
 T\kern-.1667em\lower.7ex\hbox{E}\kern-.125emX}}

\renewcommand{\nomgroup}[1]{%
 \ifthenelse{\equal{#1}{O}}{\item[\textit{Operators}]}{%
 \ifthenelse{\equal{#1}{I}}{\item[\textit{Indices}]}{%
 \ifthenelse{\equal{#1}{A}}{\item[\textit{Acronyms}]}{%
 `\ifthenelse{\equal{#1}{V}}{\item[\textit{Variables and parameters}]}{}}}}}
\usepackage{scalerel}
\usetikzlibrary{svg.path}
\definecolor{orcidlogocol}{HTML}{A6CE39}
\tikzset{
 orcidlogo/.pic={
 \fill[orcidlogocol] svg{M256,128c0,70.7-57.3,128-128,128C57.3,256,0,198.7,0,128C0,57.3,57.3,0,128,0C198.7,0,256,57.3,256,128z};
 \fill[white] svg{M86.3,186.2H70.9V79.1h15.4v48.4V186.2z}
 svg{M108.9,79.1h41.6c39.6,0,57,28.3,57,53.6c0,27.5-21.5,53.6-56.8,53.6h-41.8V79.1z M124.3,172.4h24.5c34.9,0,42.9-26.5,42.9-39.7c0-21.5-13.7-39.7-43.7-39.7h-23.7V172.4z}
 svg{M88.7,56.8c0,5.5-4.5,10.1-10.1,10.1c-5.6,0-10.1-4.6-10.1-10.1c0-5.6,4.5-10.1,10.1-10.1C84.2,46.7,88.7,51.3,88.7,56.8z};
 }
}

\newcommand\orcidicon[1]{\href{https://orcid.org/#1}{\mbox{\scalerel*{ \begin{tikzpicture}[yscale=-1,transform shape]
 \pic{orcidlogo};
 \end{tikzpicture}
 }{|}}}}

\makeatletter
\let\old@ps@headings\ps@headings
\let\old@ps@IEEEtitlepagestyle\ps@IEEEtitlepagestyle
\def\psccfooter#1{%
    \def\ps@headings{%
        \old@ps@headings%
        \def\@oddfoot{\strut\hfill#1\hfill\strut}%
        \def\@evenfoot{\strut\hfill#1\hfill\strut}%
    }%
    \def\ps@IEEEtitlepagestyle{%
        \old@ps@IEEEtitlepagestyle%
        \def\@oddfoot{\strut\hfill#1\hfill\strut}%
        \def\@evenfoot{\strut\hfill#1\hfill\strut}%
    }%
    \ps@headings%
}
\makeatother

\psccfooter{%
        \parbox{\textwidth}{\hrulefill \\ \small{24th Power Systems Computation Conference} \hfill \begin{minipage}{0.2\textwidth}\centering \vspace*{4pt} \includegraphics[scale=0.06]{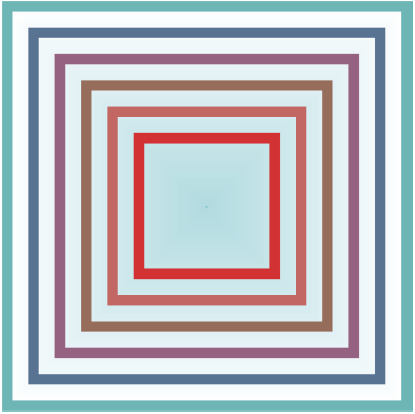}\\\small{PSCC 2026} \end{minipage} \hfill \small{Limassol, Cyprus --- June 8-12, 2026}}%
}

\begin{document}
%
\title{Multi-Region Optimal Energy Storage Arbitrage}



\author{
\IEEEauthorblockN{
Md Umar Hashmi\textsuperscript{1}, 
Harsha Nagarajan\textsuperscript{2}, 
Dirk Van Hertem\textsuperscript{1}
}

\IEEEauthorblockA{
\textsuperscript{1}KU Leuven \& EnergyVille, Leuven/Genk, Belgium, \\
\texttt{\{mdumar.hashmi, dirk.vanhertem\}@kuleuven.be}
}

\IEEEauthorblockA{
\textsuperscript{2}Applied Mathematics \& Plasma Physics (T-5), Los Alamos National Laboratory, USA, 
\texttt{harsha@lanl.gov}
}
}

\maketitle

\begin{abstract}

The increasing interconnection of power systems through AC and DC links enables energy storage units to access multiple electricity markets, yet most existing arbitrage models remain limited to single‑market participation. This gap restricts understanding of the economic value and operational constraints associated with cross‑border storage operation. To address this, an optimal multi-region energy storage arbitrage model is developed for a grid‑scale battery located at one end of an interconnector linking two distinct day‑ahead markets. The formulation incorporates battery capacity and ramping limits, converter and interconnector losses, and market‑specific buying and selling prices. 
Using disjunctive linearization of nonlinear terms, this work exactly reformulates the multi-region energy arbitrage optimization as a mixed-integer linear programming problem. The proposed formulation ensures that the battery either charges or discharges from all participating energy markets simultaneously at any given time. 
Case studies using eight years of Belgian–UK price data demonstrate that multi-region participation can increase arbitrage revenue by more than 40\% compared to local energy arbitrage operation only, while also highlighting the negative impact of interconnector congestion on achievable gains. The results indicate that cross‑border market access substantially enhances storage profitability while considering the cycle of battery and that the proposed formulation provides a computationally efficient framework for evaluating and operating storage assets in interconnected power systems.
Finally, a pseudo-efficiency term is introduced to improve battery utilization by discarding less profitable charging and discharging battery cycles.

\end{abstract}


\section*{Nomenclature}

\subsection*{Sets and Indices}
\begin{description}[leftmargin=3.5em, style=sameline]
    \item[$\bar{K}$] Mean value of vector K
    \item[$\{ \text{ch}, \text{dis}  \}$] Suffix or prefix denoting charging and discharging 
    \item[$\{A,B,..,M\}$] Unique energy markets
    \item[$i \in \{1,..,T\}$] Hourly time instances
    \item[$\lbrack k \rbrack^+, \lbrack k \rbrack^-$] Calculates $\max(0,k)$ and $\max(0,-k)$
\end{description}

\subsection*{Parameters}
\begin{description}[leftmargin=3.5em, style=sameline]
    \item[$\eta_{\text{ch}}, \eta_{\text{dis}}$] Charging and discharging efficiency
    \item[$\eta_{\text{conv}}$] Converter efficiency at both ends of interconnector
    \item[$\eta_{\text{line}}$] AC or DC interconnector line efficiency
    \item[$\zeta_i$] Interconnector rent in euros/MWh for time~$i-1$ to $i$
    \item[$b_i \in \lbrack b_{\min}, b_{\max}\rbrack$] Battery capacity for time $i$
    \item[$L^{AB}_i$] Interconnector flow from region A to B at time $i$
    \item[$L_{\max}$] Rated capacity of the interconnector
    \item[$P_i^{B,X}, P_i^{S,X}$] Buy \& sell price of electricity for region $X$,
    \item[$x_i \in \lbrack x_{\min}, x_{\max} \rbrack$] Change in battery charge level
\end{description}

\section{Introduction}
The global power network is rapidly evolving with the widespread 
adoption of AC and DC interconnections \cite{MacLeod2025}, leading to unprecedented connectivity and new business opportunities for grid-scale energy storage batteries. 
These interconnections allow batteries to participate in multiple energy markets simultaneously, significantly boosting their revenue potential compared to single-market participation. Previous use cases \cite{link2vse, link3arena, bowen2019grid, schoenung2017green, koltermann2023power} have demonstrated that grid-scale batteries are often utilized for various applications and multi-energy market products.
As energy markets become more interconnected, energy storage assets have new opportunities to maximize their operational revenue by not only participating in the local energy markets but also in other interconnected energy markets.
Since the interconnections are still substantially smaller than their local power grid, there still exist substantial inter-regional price fluctuations. Such regional differences also allow energy storage to be used for 
multi-region energy arbitrage (MREA).


Our primary objective is to develop a computationally efficient optimization model that enables grid-scale batteries to simultaneously participate in MREA. While energy storage is typically used for temporal shifting, this model achieves multi-location energy arbitrage (EA) by assuming a pseudo-agent at the other end of the interconnector.
Since the battery is physically located at one end and can only trade in the local market, pseudo-agents are employed at other interconnector ends to trade on behalf of the storage. Similar provisions exist for generators in CAISO (California) and Singapore \cite{caiso, singapore}.
The inter-regional energy transaction is possible due to an AC or DC interconnector linking two regions. Considering the limited capacity of the interconnector, MREA should also consider the cost of using the interconnector. This cost is modelled as interconnector rent in this work.
Interconnector efficiency depends on the line length, with power electronic converters incurring losses governed by technology used. Typically, interconnector losses are less than 1\% per 100 km, while each converter account for 1 \% in losses \cite{nationalgrid}. 

With the decreasing cost of grid-scale batteries due to technological advancements and improved efficiency, their widespread adoption in the power grid is inevitable \cite{hart2018energy}. NREL estimates that Li-ion batteries could be 47\% cheaper by 2030 \cite{cole2021cost}. While much literature exists on energy storage participating in various market products (e.g., energy arbitrage, ancillary services, peak shaving) {
and grid services (e.g. voltage and frequency regulation)} within a single market \cite{anderson2017co, cheng2016co, hashmi2019energy, krishnan2015optimal, lee2019closed}, this paper focuses on simultaneous participation in MREA.

There is limited research on multi-regional market participation for storage, particularly for single market products like EA with time-varying electricity prices. In \cite{hurta2022impact}, the authors observed that the separation of bidding zones between the German and Austrian day-ahead markets led to the decoupling of wholesale electricity prices and explored battery EA in both markets. Previous studies, such as \cite{bunn2010inefficient, soonee2006novel}, discussed the benefits of inter-regional energy arbitrage. For example, \cite{soonee2006novel} demonstrated that energy trade between the northern and southern grids of India, connected via HVDC back-to-back stations, resulted in savings of approximately 85 million euros from 2003 to 2005. Meanwhile, \cite{bunn2010inefficient} highlighted inefficiencies in inter-regional electricity transmission for the Anglo-French interconnector.

%
There are several notable use cases of grid-scale batteries worldwide, such as 
{
the Victorian Big Battery (300 MW/450 MWh) \cite{link2vse} and Hornsdale Power Reserve (100 MW/129 MWh) in Australia \cite{link3arena, bowen2019grid}, Moss Landing Battery (400 MW/1600 MWh) \cite{mosslanding}, Manatee Grid-Scale Battery (409 MW/900 MWh) in the USA, Edwards \& Sanborn, California (USA) (821 MW, 3,287 MWh) \cite{mortensonEdwardsSanborn}, and 
Philippsburg Battery Hub (Germany) (400 MW, 800 MWh, entering operation by 2026) \cite{enbwProjectPlan}.}
With increased penetration of distributed energy resources, there is an increased need for grid services traditionally for transmission grids \cite{aguado2017battery} and more recently for distribution networks \cite{stecca2020comprehensive}. There is a growing need for new business cases for energy storage in power networks, justifying wider investment, fuelling the energy transition.

{
The business model for energy storage would be highly influenced by the market products that the storage participates in \cite{khan2025energy, khan2026battery}. Cross-market energy arbitrage is utilized in day-ahead and intraday market products in the same region \cite{van2025optimized}.
Energy arbitrage for mobile energy storage is explored in \cite{tian2024electricity}.
The payback and investment decisions relatively depend on the price volatility of the concerned market product.
In this work, we focus on day-ahead energy arbitrage for multiple regions with two decoupled energy markets. The day-ahead prices are assumed to be known. Belgian bidding zone provides the day-ahead reference price \cite{eliaDAbid}.
For applications where prices are not known, EA needs to be coupled with price forecasting \cite{wu2025energy}.
More importantly, the proposed MREA formulation is agnostic to the market products, provided the market signals are temporally aligned in the two decoupled markets.
}

The key contributions of the paper are:

$\bullet$ \textit{Multi-region energy arbitrage (MREA) optimization}\footnote{
Energy storage arbitrage in the electricity market involves selecting optimal charging and discharging times based on time-varying electricity prices to maximize operational profit \cite{zafirakis2016value}. MREA extends this concept across multiple energy markets simultaneously.}:
{This paper proposes} an optimal Mixed-Integer Linear Programming (MILP) formulation for simultaneous participation of energy storage in energy arbitrage across \textit{multiple} energy markets. The novel formulation considers multiple energy markets' price fluctuations, interconnector capacity limitations and limiting battery cycles while ensuring it is not charging and discharging simultaneously.

$\bullet$ \textit{Energy Storage Profitability Assessment}: A comprehensive framework for the techno-economic evaluation of grid-scale energy storage performing MREA {is provided}. This assessment considers the battery's operational and shelf life limitations while projecting annual simulated revenue. 8 years of MREA simulation using real day-ahead electricity prices in Belgium and the UK is performed for \href{https://www.nemolink.co.uk/}{NEMO Link} interconnector linking Belgian and British power networks. 
{Furthermore, the formulation accounts for battery health while improving storage utilization by integrating a tunable pseudo-efficiency term to discard excessive storage-device cycles, thereby enhancing per-cycle revenue for the storage.}

$\bullet$ \textit{Benchmarking energy arbitrage models}: 
{
The MREA formulation is compared with classical LP, MILP, and heuristic no‑discharge models used for single‑market arbitrage, especially under conditions of negative electricity prices. The comparison highlights the revenue improvements and utilization benefits enabled by multi‑region market participation.
Finally, the proposed model for MREA and single-region MILP models is observed to be performing well for negative electricity prices, where the traditional energy arbitrage formulations are no longer convex and computationally reliable. 
}


{
This paper is organized in sections.
Section \ref{section2} summarizes the energy arbitrage models while considering negative electricity prices.
Section \ref{section3} details the proposed MREA model considering two decoupled grids (independent energy markets) connected via an interconnector.
Numerical case studies are presented in Section \ref{section4}.
Section \ref{section5} concludes the paper.
}

\section{Single Region Energy arbitrage approximates}
\label{section2}

{
In this section, the battery model is presented that will be used for single and multi-region EA.
The state-of-the-art single-region EA models are detailed, and their ability to handle negative electricity prices is discussed.
The single-region EA arbitrage models detailed in this section are benchmarked with the MREA model proposed in Section \ref{section3}.
}

\subsection{Battery model}
The battery model incorporates ramping and capacity constraints, along with charging and discharging efficiencies, denoted by $\eta_{\text{ch}}, \eta_{\text{dis}} \in (0,1]$. The energy optimization considers the change in battery energy levels at time $i$, denoted as $x_i$. This change is defined as $x_i = h \delta_i$, where $\delta_i \in [\delta_{\min}, \delta_{\max}]$ represents the battery's ramp, and $h$ is the sampling period. $\delta_i > 0$ when charging, and $\delta_i < 0$ when discharging, with $\delta_i$ in units of power (MW) and $x_i$ in energy (MWh). The battery charge is given by:
\begin{equation}
b_i = b_{i-1} + x_i, \quad b_i\in [b_{\min},b_{\max}], \forall i,
\label{eq:cap}
\end{equation}
where $b_{\min}$ and $b_{\max}$ are the minimum and maximum battery capacities. The power consumed by the battery at time $i$ is:
\begin{equation}
f(x_i)= \frac{[x_i]^+}{h \eta_{\text{ch}}} - \frac{\eta_{\text{dis}}[x_i]^-}{h}=\frac{\max(0,x_i)}{h\eta_{\text{ch}}} - \frac{\eta_{\text{dis}}\max(0,-x_i)}{h}, 
\label{finverse}
\end{equation}
where $x_i$ lies within the range $X_{\min} = \delta_{\min} h$ to $X_{\max} = \delta_{\max} h$. 
The ramping constraint is given as:
\begin{equation}
    x_i \in [X_{\min}, X_{\max}].
    \label{eq:ramp}
\end{equation}

The battery is interfaced via a converter (inverter/rectifier) with efficiency $\eta_{\text{conv}} \in (0,1]$. The modified battery charging and discharging efficiency is denoted as:
\begin{equation}
    \eta_{\text{ch}}^* = \eta_{\text{ch}}\cdot\eta_{\text{conv}},~~~~
    \eta_{\text{dis}}^* = \eta_{\text{dis}}\cdot\eta_{\text{conv}}.
    \label{eq:eff}
\end{equation}


The modified battery efficiency in \eqref{eq:eff} includes converter efficiency. For HVDC interconnections, converter efficiency is about {98}\% (considering converters at both ends) \cite{wikihvdc}, making it essential to account for these losses.

\subsection{Energy storage arbitrage models}

Energy storage arbitrage has emerged as a critical application for enhancing grid stability and capitalizing on the volatility introduced by renewable energy sources, fundamentally involving the strategy of purchasing electricity at low prices to charge a storage system and selling it back to the grid when prices are high. To maximize profitability, operators employ sophisticated optimization models to schedule the charging and discharging cycles. A fundamental approach is to frame the problem as a convex \cite{hashmi2017optimal} or linear program (LP) \cite{hashmi2019optimal,he2015optimal} or
MILP \cite{mercier2023value, shen2020modeling} or dynamic programming (DP) \cite{cheng2016co, zheng2022arbitraging}
where the goal is to maximize revenue over a time horizon \(T\). 

\subsection{Dealing with negative electricity prices}
\begin{figure}[!htbp]
	\center
	\includegraphics[width=3.4in]{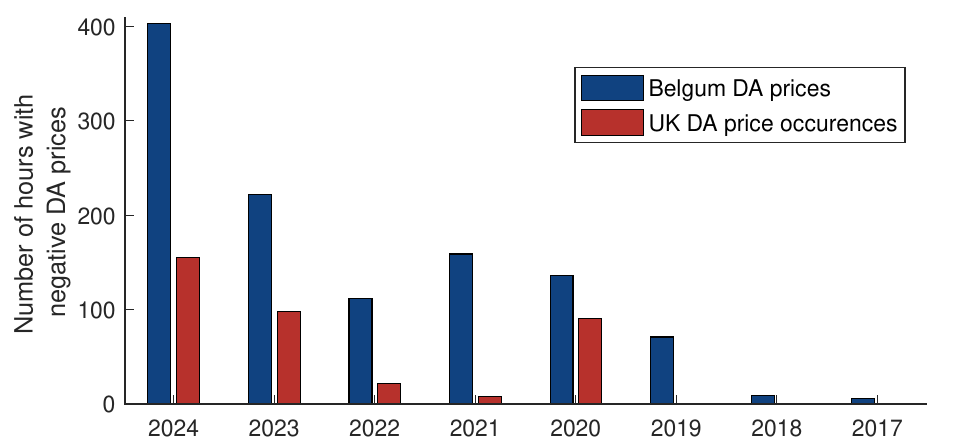}
	\vspace{-2pt}
	\caption{\small{Occurrences of negative prices in the DA market in Belgium and the UK for the last 8 years \cite{entsoeTransparencyPlatform}.}}
	\label{case2fig}
\end{figure}

The electricity price in day-ahead (DA) and intraday electricity markets is increasingly becoming negative. Figure \ref{case2fig} shows the number of hours with negative DA prices in Belgium and the UK in the last years. Note the growing trend in such occurrences. 

The classical formulation of the EA problem relies on the assumption of non-negative electricity prices, which ensures the problem is convex and solvable via methods like LP. This assumption is now frequently violated in markets with high renewable penetration. The presence of negative prices introduces a non-convexity into the optimization, as profit can be generated from two distinct actions: selling high or being paid to charge low. To sidestep this mathematical challenge, some proposed models employ a restrictive heuristic such as the \textit{noDis} model, where charging is only permitted when prices are negative (i.e. no discharging permitted) \cite{wu2025energy}. 
This simplification, however, leads to suboptimal outcomes, as it arbitrarily foregoes profitable cycles between relative negative price fluctuation values.

Although the convexity of the EA problem cannot be ensured with negative electricity prices. LP, MILP, \textit{noDis}  formulations can be compared with the DP model for EA, as proposed in \cite{cheng2016co}. As the decision variable discretization is made smaller, the DP-based formulation will converge to an optimal solution, but solving this is computationally hard due to the curse of dimensionality \cite{powell2007approximate}.

{
\subsection{Mathematical model for single region EA}
EA is performed for time-varying electricity prices denoted as $P^B_i$ and $P^S_i$ representing buying and selling prices. The EA objective function is denoted as
\begin{equation}
C_i=
\begin{cases}
P^B_i . x_i / \eta_{\text{ch}} ,& x_i \geq 0 ,\\
P^S_i . x_i . \eta_{\text{dis}}, & x_i < 0
\end{cases}
\end{equation} 
%
%
\subsubsection{Linear formulation}
The LP model for a single-region EA is proposed in \cite{hashmi2019optimal} and given as below
\begin{topbot2}
    \text{EA with linear formulation ($P_{\text{LP}}$)}
\end{topbot2}
\vspace{-10pt}
\begin{IEEEeqnarray}{ l C l }
    \label{eq:original_formulation1}
    \text{Obj. func.:~~} 
     &\min{ \quad \{t_1 + t_2+...+t_N\}},&  \\
    \text{subject to:} & \eqref{eq:cap}, \eqref{eq:ramp} & \nonumber \\
    & {P^B_i} x_i /\eta_{\text{ch}} \leq t_i, ~\forall i & \label{segment1} \\
    &{P^S_i}{\eta_{\text{dis}}} x_i  \leq  t_i, ~\forall i& \label{segment2}\\
    \hline \nonumber
\end{IEEEeqnarray}
where $t_i$ denote the epigraph at time $i$. This formulation is possible due to a piecewise linear objective function.
This model is extended to consider storage ramp‑rate limits in \cite{hashmi2024linear}.
\subsubsection{MILP formulation}
\label{sec:SREAmilp}
The MILP formulation for EA 
maximizes the objective function \(\sum_{i \in T} (P^S_i.x_i^{dis}\eta_{dis} - P^B_i.x_i^{ch}/\eta_{ch})\), where \(\ P^S_i, P^B_i\) are the sell and buy electricity price, and \(x_i^{ch}\) and \(x_i^{dis}\) are the charging and discharging power at time \(i\), respectively. This optimization is subject to linear constraints governing the system's physics, such as the state of charge dynamics \(b_i = b_{i-1} + x_i^{ch} - x_i^{dis}\), and limits on power and energy capacity. For greater operational fidelity, this model is often extended to an MILP to incorporate discrete decisions \cite{mercier2023value}. By introducing binary variables, \(z_i^{ch}\) and \(z_i^{dis}\), an MILP can enforce mutually exclusive operating states through constraints like \(z_i^{ch} + z_i^{dis} \leq 1\), preventing simultaneous charging and discharging, thereby providing a more realistic and robust scheduling tool \cite{sang2022electricity}.\\
An alternate MILP formulation without separating charging and discharging variables, adapted for the epigraph formulation, can be presented as follows
\begin{topbot2}
    \text{EA with MILP formulation ($P_{\text{MILP}}$)}
\end{topbot2}
\vspace{-10pt}
\begin{IEEEeqnarray}{ l C l }
    \label{eq:original_formulation2}
    \text{Obj. func.:~~} 
     &\min{ \quad \{t_1 + t_2+...+t_N\}},&  \\
    \text{subject to:} & \eqref{eq:cap}, \eqref{segment1}, \eqref{segment2} & \nonumber \\
    & x_i \in [(1-z_i).X_{\min}, z_i.X_{\max}], z_i \in \{0,1\}& \label{milp_eq} \\
    \hline \nonumber
\end{IEEEeqnarray}
\subsubsection{noDis formulation}In this model, based on \cite{wu2025energy}, the battery can only charge for negative prices. This is implemented by redefining the ramping constraint as
\begin{equation}
    x_i \in [\xi_i.X_{\min}, X_{\max}],
    \label{eq:ramp_noDis}
\end{equation}
where $\xi_i$ equals 1 for $P^B_i < 0$ and 0 otherwise. 
For the $noDis$ model, $P_{\text{LP}}$ is solved for updated ramp constraint \eqref{eq:ramp_noDis}.
\subsubsection{Dynamic programming formulation}
The DP EA benchmarking model used in this work is based on prior works \cite{cheng2016co, zheng2022arbitraging}.
Denote the action grid using discrete steps $d_A$ and $d_S$ as
\begin{equation}
    \chi =\{ X_{\min}: d_A:  X_{\max}   \}, \quad 
    \Psi =\{ b_{\min}: d_S : b_{\max}   \},
\end{equation}
\begin{topbot2}
    \text{EA with DP formulation ($P_{\text{DP}}$)}
\end{topbot2}
\vspace{-10pt}
\begin{IEEEeqnarray}{ l l }
    \label{eq:original_formulation3}
     x_i^* \text{=} \arg \min_{\chi, \Psi} [ C_i + V_{i+1}(b_i+x_i)],  i \in \{N, N-1,..,1\}&  \\
      \text{subject to: }\eqref{eq:cap}, \eqref{eq:ramp}, \text{ initial charge }b_0  & \nonumber \\
    \hline \nonumber
\end{IEEEeqnarray}
As the discrete step size $d_S$ and $d_A$ decrease, the state and action grids explode in size. Backward induction must evaluate every action at every state for each time step, so the runtime scales with the product of horizon length, number of SOC points, and number of actions. Finer steps also increase interpolation calls, bounds checks, and branching, amplifying instruction count and cache misses. Memory use grows for the value table and policy matrix. 
}

\section{Optimal multi-regional arbitrage}
\label{section3}

Given the grid configuration, a battery can be strategically located at one end of the interconnector to capitalize on pricing disparities between the two (or potentially more than two) grids over charge and discharge cycles. In this study, the battery is assumed to be a price-taker in the electricity market.

\subsection{Multi-regional energy arbitrage model}
\begin{figure*}[!htbp]
	\center
	\includegraphics[width=5.58in]{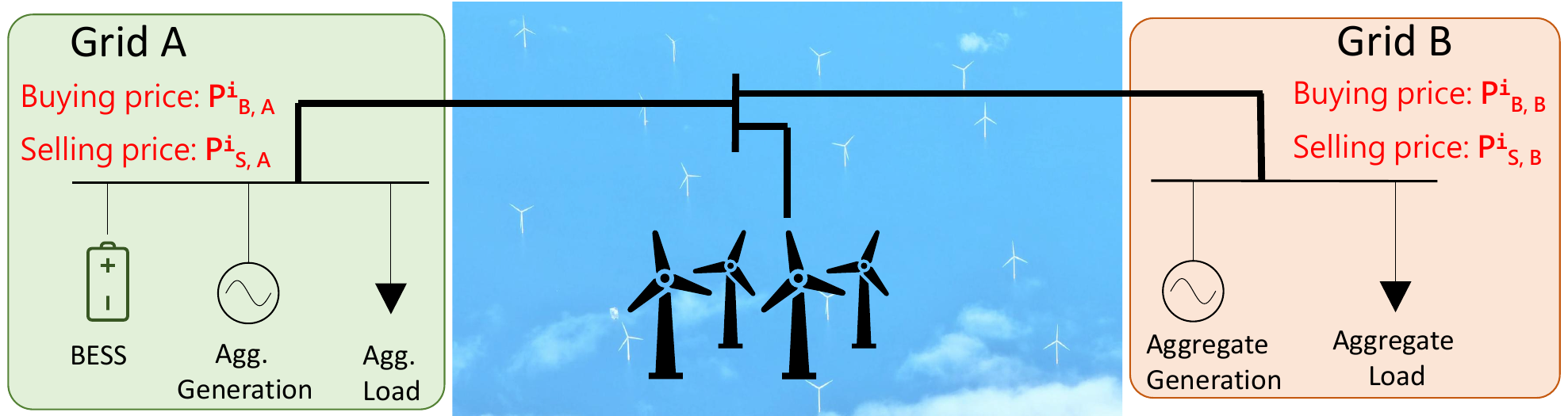}
	\vspace{2pt}
	\caption{\small{Stylized grids connected via an AC or DC interconnector with an offshore wind injection located in the centre.}}
	\label{topo_fig}
\end{figure*}

{This subsection details the formulation for}
an optimal energy arbitrage model for energy storage participating in MREA between two regions, referred to as grids A and B in Figure \ref{topo_fig}. Since grids A and B have separate energy markets, the price levels for consumption and injection vary due to the limited size of the interconnector relative to the cumulative power needs of both grids\footnote{Interconnectors operated under a flow-based market reduce price fluctuations, making inter-regional energy trades {relatively} less profitable as they grow \cite{leuven2015cross}.}. 

In grid A, $P_i^{\text{\sc b,a}}$ and $P_i^{\text{\sc s,a}}$ denote the buying (consumption) and selling (injection) prices of electricity at time $i$, respectively. Similarly, in grid B, $P_i^{\text{\sc b,b}}$ and $P_i^{\text{\sc s,b}}$ denote the buying and selling prices at time $i$.
As shown in Figure \ref{topo_fig}, the battery is located in grid A. The efficiency of the AC or DC interconnector linking grids A and B is $\eta_{\text{line}} \in (0,1]$.
The energy storage and interconnector are assumed to be owned by different entities, with the interconnector owner charging a fee, $\zeta_i > 0, \forall i$, for energy transferred. 
The total interconnector rent for time $i$ is proportional to the energy imported or exported via the interconnector between intervals $i-1$ and $i$.

The effective buy and sell prices in grid~A for transactions settled in grid~B (with interconnector rent and losses) are
\begin{subequations}\label{eq:effective_prices}
\begin{align}
\widetilde{P}_i^{\text{\sc b,b}} &= \frac{P_i^{\text{\sc b,b}}+\zeta_i}{\eta_{\text{line}}}, \label{eqbuypriceredef}\\
\widetilde{P}_i^{\text{\sc s,b}} &= \left(P_i^{\text{\sc s,b}}-\zeta_i\right)\eta_{\text{line}}. \label{eqsellpriceredef}
\end{align}
\end{subequations}
Interconnector losses and rent therefore increase the effective buying price in grid~A and reduce the effective selling price.

\begin{figure*}[!htbp]
	\center
	\includegraphics[width=5.46in]{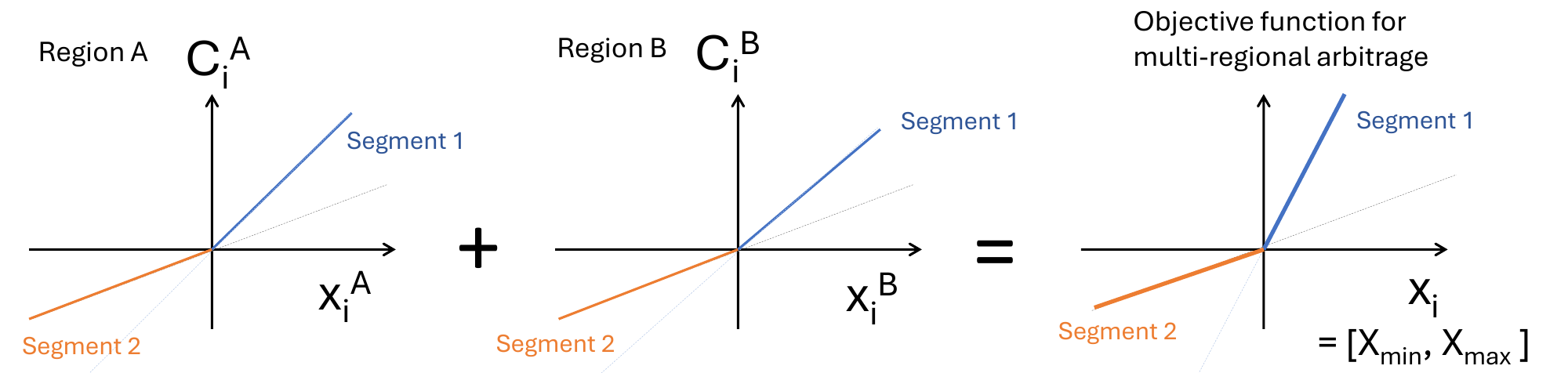}
	\vspace{2pt}
	\caption{\small{Piecewise objective function for MREA optimization.
 }}
	\label{fig:objfunc}
\end{figure*}

\textbf{Multi-region energy arbitrage cost function}:
The battery operating costs for transactions in grids A and B are given by:
\begin{equation}
    C_i^{\text{\sc a}} = P_i^{\text{\sc b,a}} \frac{[x_{i}^{\text{ \sc a}}]^+}{\eta_{\text{ch}}^*} - P_i^{\text{\sc s,a}} [x_{i}^{\text{ \sc a}}]^- \eta^*_{\text{dis}},
    \label{eq:objfun1}
\end{equation}
\vspace{-10pt}
\begin{equation}
    C_i^{\text{\sc b}} = \widetilde{P}_{i}^{\text{\sc b,b}} \frac{[x_{i}^{\text{ \sc b}}]^+}{\eta^*_{\text{ch}}} - \widetilde{P}_{i}^{\text{\sc s,b}} [x_{i}^{\text{ \sc b}}]^- \eta^*_{\text{dis}}, 
    \label{eq:objfun2}
\end{equation}
where $x_{i}^{\text{ \sc a}}, x_{i}^{\text{ \sc b}} \in [X_{\min}, X_{\max}]$ denote the change in battery charge level due to grid A and B, respectively.
{Note that the piecewise linear objective function given in \eqref{eq:objfun1} and \eqref{eq:objfun2}, is convex if and only if the slope of the cost function is monotonically increasing. 
This is ensured if
$P_i^{\text{\sc b,a}}/\eta_{\text{ch}}^* \geq P_i^{\text{\sc s,a}} \eta^*_{\text{dis}} ~ \forall i$ for \eqref{eq:objfun1} and
$\widetilde{P}_{i}^{\text{\sc b,b}}/ \eta^*_{\text{ch}} \geq \widetilde{P}_{i}^{\text{\sc s,b}}  \eta^*_{\text{dis}} \forall i$ for \eqref{eq:objfun2}.
}
The cost functions for MREA optimization are illustrated in Figure \ref{fig:objfunc}. As shown, each cost function consists of two linear segments, with extrapolation indicated by a dotted line.

The battery ramping constraint, considering participation from grids A and B, must comply with the overall battery ramping requirements. The constraint is defined as:
\begin{equation}
    (x_{i}^{\text{ \sc a}} + x_{i}^{\text{ \sc b}}) \in [X_{\min}, X_{\max}].
    \label{eq:ramp2a}
\end{equation}
Note that $x_{i}^{\text{\sc a}}$ and $x_{i}^{\text{\sc b}}$ must share the same sign, so the battery either charges or discharges in a given interval, but not both. This is enforced by the constraint:
 \begin{equation}
     x_{i}^{\text{ \sc a}} \cdot x_{i}^{\text{ \sc b}} \geq 0 \quad \forall i.
     \label{eqbilinear}
 \end{equation}
As it is, constraint \eqref{eqbilinear} is intractable due to the bilinear term and its non-convex feasible region. Conventional methods like McCormick relaxation \cite{mccormick1976computability} are unsuitable, as they can lead to weak relaxations and impractical solutions, allowing simultaneous charging and discharging. Instead, we choose to exactly reformulate the nonlinear constraint \eqref{eqbilinear} into a set of linear constraints, based on the following key observation: For each time period $i$, the feasible region in the space of variables $(x_{i}^{\text{\sc a}}, x_{i}^{\text{\sc b}})$ is either the non-negative or the non-positive orthant. Since this feasible region is piecewise convex \cite{nagarajan2019adaptive, yang2022optimal}, we can reformulate it to represent the disjunctive union of these two regions by introducing two binary variables $z_i^{\text{ch}}, z_i^{\text{dis}} \in \{0,1\}$ and satisfying the following inequalities $\forall i$:
 \begin{subequations}
 \begin{align}
     & x_{i}^{\text{ \sc a}} \geqslant z_i^{\text{ch}} \cdot X_{\min}, \quad x_{i}^{\text{ \sc a}} \leqslant z_i^{\text{dis}} \cdot X_{\max},\\
     & x_{i}^{\text{ \sc b}} \geqslant z_i^{ch} \cdot X_{\min}, \quad x_{i}^{\text{ \sc b}} \leqslant z_i^{\text{dis}} \cdot X_{\max},\\
     & z_i^{\text{ch}} + z_i^{\text{dis}} = 1.0. 
 \end{align}
 \label{eq:linearization}
 \end{subequations}
{
Variable splitting with disjunctive binary modeling is standard in single-region energy arbitrage. In that setting, the battery power (or energy change) is typically represented as $x_i = x_i^{\text{ch}} - x_i^{\text{dis}}$ (see Sec.~\ref{sec:SREAmilp}), with a binary variable enforcing that at most one of $x_i^{\text{ch}}$ and $x_i^{\text{dis}}$ is nonzero, thereby preventing simultaneous charging and discharging.}

{The disjunctive binary model in \eqref{eqbilinear}--\eqref{eq:linearization} is different. In the proposed MREA formulation, the binary variable enforces a \emph{shared} operating mode across regions, so the battery either charges or discharges across all regions within an interval, avoiding unrealistic simultaneous charging in one market and discharging in another. Importantly, we achieve this without explicitly introducing separate charge/discharge variables for each region. This distinction becomes critical for scalability: a direct extension of single-region splitting with binaries does not scale well to $M$ regions with independent markets.
}

\subsection{Epigraph-based mathematical formulation}

The MREA problem is formulated as a constrained minimization of a convex piecewise-linear objective subject to linear constraints and binary variables:

\begin{topbot2}
    \text{Original problem ($P_{\text{orig}}$)}
\end{topbot2}
\vspace{-10pt}
\begin{IEEEeqnarray}{ l C l }
    \label{eq:original_formulation4}
    \text{Objective function:~~} 
     &\min{ \sum_i^N (C_i^{\text{\sc a}} ~ + ~C_i^{\text{\sc b}}) },& \label{eqoriOBJ} \\
    \text{subject to:} & \eqref{eq:cap}, \eqref{eq:ramp2a}, \eqref{eqbilinear} & \nonumber \\
    \hline \nonumber
\end{IEEEeqnarray}

$P_{\text{orig}}$ can be equivalently reformulated as an MILP via an epigraph transformation \cite{hashmi2019optimal}, yielding:

\begin{topbot2}
    \text{Epigraph formulation for MREA ($P_{\text{MILP}}^{\text{MR}}$)}
\end{topbot2}
    \vspace{-8pt}
{\allowdisplaybreaks
\begin{IEEEeqnarray}{ l C l }
    \label{eq:epi_formulation}
    \text{Objective function:} &~& \nonumber \\
    \min_{x_{i}^{\text{ \sc a}}, x_{i}^{\text{ \sc b}}}{ \quad \sum_{i=1}^{N}\Big(t_{i}^{\text{\sc a}}+t_{i}^{\text{\sc b}}\Big) }, &~& \label{eqoriOBJ_epi} \\
    \text{subject to:} && \nonumber \\
    \text{Cost function segment 1 and 2:} &~& \nonumber \\
    P_i^{\text{\sc b,a}} \frac{x_{i}^{\text{ \sc a}}}{\eta^*_{\text{ch}}} \leq t_{i}^{\text{ \sc a}}, &~& \forall~~ i,
    \label{eq14}\\
    P_i^{\text{\sc s,a}} x_{i}^{\text{ \sc a}} \eta^*_{\text{dis}} \leq t_{i}^{\text{ \sc a}}, &~& \forall~~ i, \label{eq15}\\
    \text{Cost function segment 3 and 4:} &~& \nonumber \\
    \widetilde{P}_{i}^{\text{\sc b,b}} \frac{x_{i}^{\text{ \sc b}}}{\eta^*_{\text{ch}}} \leq t_{i}^{\text{ \sc b}}, &~& \forall~~ i, \label{eq16}\\
    \widetilde{P}_{i}^{\text{\sc s,b}} x_{i}^{\text{ \sc b}} \eta^*_{\text{dis}} \leq t_{i}^{\text{ \sc b}}, &~& \forall~~ i, \label{eq17}\\   
 \text{Ramp constraint:~} &~& \eqref{eq:ramp2a} \nonumber \\
 x_{i}^{\text{ \sc a}} \in [X_{\min}, X_{\max}],&~& ~\forall~ i,\\
 x_{i}^{\text{ \sc b}} \in [X_{\min}, X_{\max}],&~& ~\forall~ i, 
 \label{eq:rampxB} \\
 \text{Capacity constraint:~} &~& \nonumber \\
 \text{$\sum$} \{x_{j}^{\text{ \sc a}} + x_{j}^{\text{ \sc b}}\} \leq b_{\max}-b_0,~&~& \forall~ j \in\{1,.. i\}, \label{eq20} \label{eqcap1}\\ \vspace{-5pt}
 - \text{$\sum$} \{x_{j}^{\text{ \sc a}} + x_{j}^{\text{ \sc b}}\} \leq b_0- b_{\min},~&~& \forall~ j \in\{1,.. i\}, \label{eq21} \label{eqcap2} \\
 \text{Constraints} \ \eqref{eq:linearization} \ \text{to ensure:~} x_{i}^{\text{ \sc a}}\cdot x_{i}^{\text{ \sc b}}\geq 0, &~& \forall i   \\
    \hline \nonumber
\end{IEEEeqnarray}
}

Reformulating the problem in this way is useful because $P_{\text{MILP}}^{\text{MR}}$ introduces separate epigraph variables for each region at each time step, thereby decoupling the piecewise-linear cost terms while coupling the markets only through the battery’s operating constraints. The formulation extends naturally to more than two markets by enforcing (i) the aggregate throughput constraint
$$
x_i^{A}+x_i^{B}+\cdots+x_i^{M}\in [X_{\min},X_{\max}],
$$
and (ii) a shared charge/discharge mode across all participating markets by applying \eqref{eq:linearization} to each regional variable $x_i^{m}$ using the same binaries $z_i^{\text{ch}},z_i^{\text{dis}}$ (i.e.,  $\forall \ i, \ x_i^{A},x_i^{B},\ldots,x_i^{M}$ lie in the same orthant). In this work, we focus on participation in only two energy markets.

\subsection{Impact of interconnector flow}
AC or DC interconnections linking two separate energy markets have a limited capacity, denoted as $L_{\max}$. The MREA formulation, $P_{\text{MILP}}^{\text{MR}}$, assumes the interconnector does not limit battery charging and discharging, as shown in \eqref{eq:rampxB}. This implies $\max(|X_{\min}|,X_{\max}) \leq L_{\max}$, with interconnector capacity available at all times. However, this assumption may not hold, as interconnectors are often used for other purposes, such as renewable energy curtailment reduction, security of supply, energy trade, and ancillary services \cite{kaushal2019overview, maciver2021electrical, trovato2021flexible, cassarino2022meeting}, and may not be fully available.

We denote the flow in the interconnector (such that the battery is located at grid A) as $L^{\text{\sc ab}}_i$, where $L^{\text{\sc ab}}_i > 0$ indicates power injected from {region A to B} and vice versa. The interconnector flow may affect the battery's ability to charge or discharge.
To account for this, \eqref{eq:rampxB} ({ramping constraint for region B}) needs to be updated, considering the interconnector flow. The operating envelopes are defined as follows:
\begin{equation}
        X_{\max}^{\text{adj}} = \begin{cases}
                       \max\big(0, \min(X_{\max}, L_{\max} + L^{\text{\sc ab}}_i)\big), \text{~if $L^{\text{\sc ab}}_i < 0$}, \\
                        X_{\max}, \text{~if $L^{\text{\sc ab}}_i \geq 0$}.
                    \end{cases}
         \label{eqlimmax1}           
\end{equation}
\vspace{-10pt}
\begin{equation}
        X_{\min}^{\text{adj}} = \begin{cases}
                       X_{\min}, \text{~if $L^{\text{\sc ab}}_i < 0$}, \\
                       \min\big(0, \max(X_{\min}, -L_{\max} + L^{\text{\sc ab}}_i)\big), \text{~if $L^{\text{\sc ab}}_i \geq 0$}.
                    \end{cases}
         \label{eqlimmax2}           
\end{equation}
where \eqref{eqlimmax1} and \eqref{eqlimmax2} show that due to interconnector flows, the availability for charging or discharging the storage via the other end of the interconnector may not hold. For the optimization problem $P_{\text{MILP}}^{\text{MR}}$, this is addressed by adjusting the ramping constraint, modifying \eqref{eq:rampxB} accordingly:
\begin{equation}
     x_{i}^{\text{ \sc b}} \in [X_{\min}^{\text{adj}}, X_{\max}^{\text{adj}}], ~\forall~ i.
     \label{rampup1}
\end{equation}

\subsection{Reducing cycles of operation}
The cycles of operation can be reduced by limiting the lower transaction charging and discharging cycles of MREA. This is achieved by introducing a pseudo efficiency term, denoted as $\eta_{\text{pseudo}}$.
This is modelled by modifying \eqref{eq:eff} as 
\begin{equation}
    \eta_{\text{ch}}^* = \eta_{\text{ch}}\cdot\eta_{\text{conv}}\cdot \eta_{\text{pseudo}},~~~~
    \eta_{\text{dis}}^* = \eta_{\text{dis}}\cdot\eta_{\text{conv}}\cdot \eta_{\text{pseudo}}.
    \label{eq:eff2}
\end{equation}


\subsection{{Embedding uncertainty in the MREA model}}
{The proposed MREA formulation is deterministic, consistent with common practice where storage operators optimize against published DA prices. The model can be extended to account for uncertainty as follows.}


{
\subsubsection{When DA electricity prices are uncertain}
We embed price uncertainty via a sample average approximation (SAA) of the expected operating cost. Specifically, given $S$ (equiprobable) price/initial-SOC scenarios, we replace \eqref{eqoriOBJ_epi} with
\begin{equation}
\min \quad \frac{1}{S}\sum_{s=1}^{S}\sum_{i=1}^{N}\Big(t_{i,s}^{\text{\sc a}}+t_{i,s}^{\text{\sc b}}\Big),
\end{equation}
where $s\in\{1,\ldots,S\}$ and $t_{i,s}^{\text{\sc a}},t_{i,s}^{\text{\sc b}}$ are the scenario-specific epigraph variables induced by scenario-$s$ prices. If a DA schedule must be fixed before prices realize, the dispatch variables are shared across scenarios (nonanticipativity); otherwise, recourse decisions can be indexed by $s$.
}

{
\subsubsection{Near real-time MREA implementation}
This can be handled using a model predictive control (MPC) implementation: at each time step, solve the (deterministic or scenario-based) MREA over a rolling horizon (e.g., 6--48~h, depending on forecast fidelity) and implement only the first-period decision \cite{hashmi2019energy}. Price forecasting is beyond the scope of this work.
}

\section{Numerical case study}
\label{section4}

\begin{figure}[!htbp]
	\center
	\includegraphics[width=3.3in]{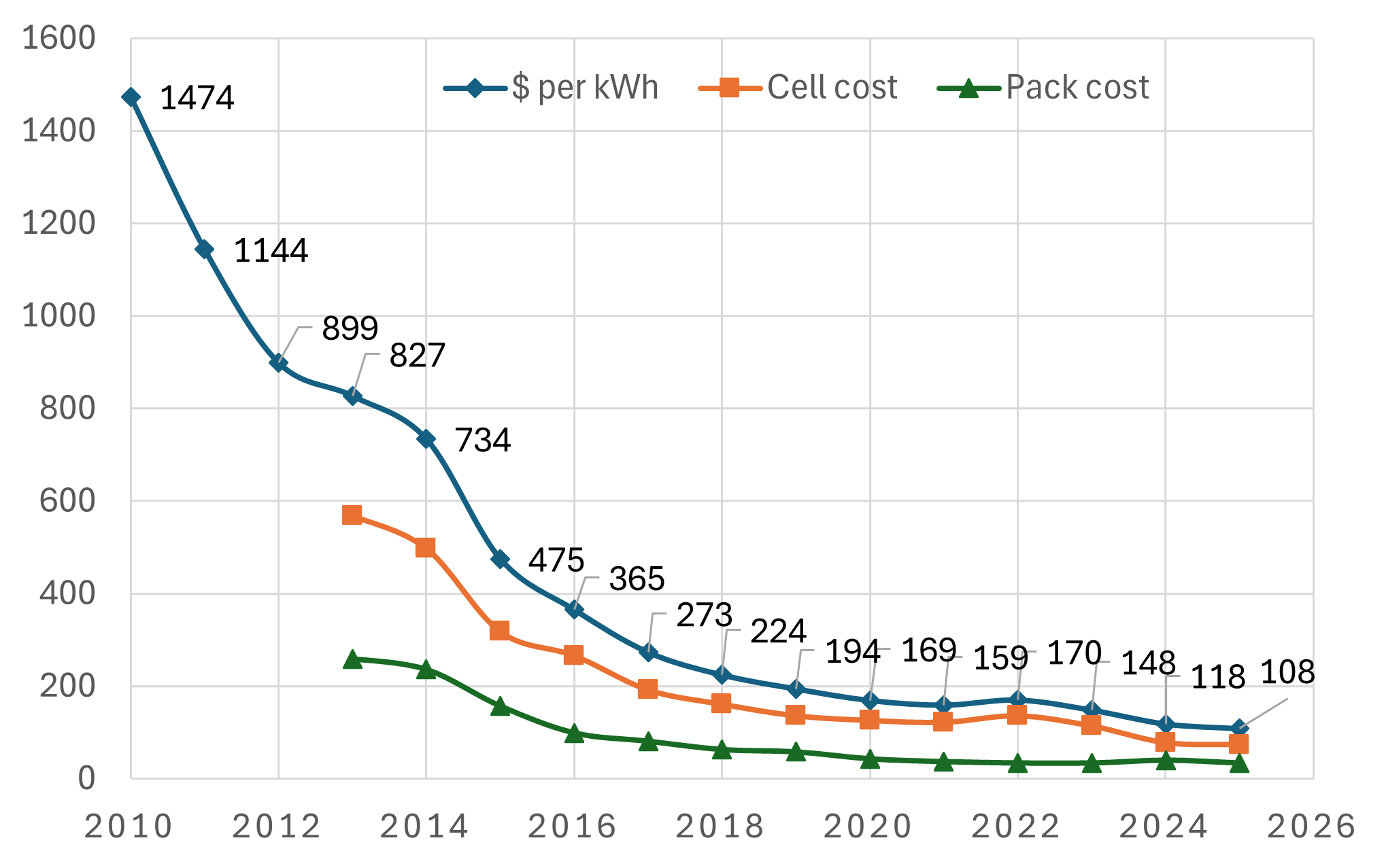}
	\vspace{-2pt}
	\caption{\small{{Battery prices (per kWh) based on BloombergNEF’s 2025 survey show average lithium‑ion pack costs dropping 8\% year‑on‑year, extending the decline of 93\% since 2010 \cite{bnefLithiumIonBattery}.}}}
	\label{bnefCost}
\end{figure}

The numerical case studies are based on 
battery characteristics detailed in Table \ref{tab:batparameters}.
{
For a reasonable cost-benefit analysis, the storage cost is selected based on BloombergNEF’s 2025 Lithium‑Ion Battery Price Survey, as shown in Figure \ref{bnefCost} \cite{bnefLithiumIonBattery}.
The battery cost assumed in this work is 100 euros/kWh (after currency adjustment for \$1.08 per euro).
The battery size is selected as 1 MW compared to 1 GW NEMO link capacity, as the interconnector has many objectives, such as (i) load balancing, (ii) network security, (iii) accommodating a large share of RES by reducing curtailment, (iv) profit maximisation for the stakeholders, etc.
Due to these multiple objectives, the interconnector normally operates at full capacity, leaving behind small headroom for a large storage to perform MREA.
Based on the interconnector flow data, it is observed that 1 MW battery could provide a compelling business case for MREA. 
Also, for the case where the interconnector capacity is not saturated, the net MREA revenue of a bigger battery can be scaled linearly for the same C-rate (0.5C-0.5C\footnote{The notation xC-yC expresses how quickly a battery can be charged and discharged. 
} in this case).
Further work is needed for optimal sizing of the battery, which goes beyond the scope of this work.
}

\begin{table}[!htbp]
\centering
\caption{\small{Battery parameter used for simulations}}
\footnotesize
\begin{tabular}{l|l} 
\hline
Attributes & Value \\
\hline \hline
Capital cost                       & 100 euros/kWh  \\ 
Rated capacity ($b_{\max}$)              & 1 MWh          \\ 
Minimum operational capacity ($b_{\min}$)  & 0.1 MWh        \\ 
Max charging rate    ($\delta_{\max}$)         & 0.5 MW         \\ 
Min discharging rate   ($\delta_{\min}$)      & -0.5 MW        \\ 
Charging efficiency      ($\eta_{\text{ch}}$)    & 0.95           \\ 
Discharging efficiency ($\eta_{\text{dis}}$)      & 0.95           \\ 
Converter efficiency    ($\eta_{\text{conv}}$)     & 0.95           \\ 
Initial charge level  ($b_0$)       & 0.5 MWh        \\ 
Cycle life (100\% DoD)       & 7200           \\ 
Calendar life                & 10 years       \\
\hline
\end{tabular}
\label{tab:batparameters}
\end{table}

The case studies are performed for a battery connected at one end of the NEMO link connecting mainland Belgium to the UK grid via a 140 km undersea cable with a 1 GW capacity and 2.5\% losses \cite{linknemoloss}.
Note that the currencies in Belgium and the UK are different; for this work, the British pound is converted to euros with a factor of 1.15.

\subsection{Performance indices}
The performance indices used for evaluating numerical case studies are as follows:


\textbf{Revenue from arbitrage}
The revenue from performing EA in multiple markets is denoted as
$
    R = -\sum_i^N \{C_i^{\text{\sc a}} ~ + ~C_i^{\text{\sc b}}\}.
$

\textbf{Cycles of operation of the battery}:
Battery life is often defined in terms of cycle life and calendar life. Cycle life is determined by the number of charging and discharging cycles, with a non-linear relationship between the cycle and depth of discharge. In this work, {Algorithm 1 from \cite{hashmi2018long} is used} to calculate the cycles of operation, based on the yearly charge and discharge trajectory from solving $P_{\text{MILP}}^{\text{MR}}$. Accurate cycle counting is essential for assessing the financial viability of battery installation.



%

\textbf{Computation time}:
The simulations are performed on a laptop with an HP Intel(R) Core(TM) i7 CPU at 1.90 GHz and 32 GB RAM, using Matlab 2021a.
{
The Matlab's \texttt{intlinprog} \cite{intlinprog} solver utilizes the default ``HiGHS'' algorithm \cite{highsSolver}.
}

\textbf{Revenue per cycle}: {is the 
ratio of revenue and the cycles of operation}. This index provides insights into the utilization quality of the battery. Higher revenue per cycle is desired for maximizing the revenue and battery operational life.

\textbf{Verifying charge discharge conflict}: As mentioned previously, the battery cannot simultaneously charge or discharge from regions A and B. In order to ensure this, the following metric {is assessed}:
$$M_{\text{ind}} = \max_i~ x_{i}^{\text{ \sc a}} \cdot x_{i}^{\text{ \sc b}}$$

\subsection{Case study 1: Benchmarking arbitrage models}
\begin{figure}[!htbp]
	\center
	\includegraphics[width=3.4in]{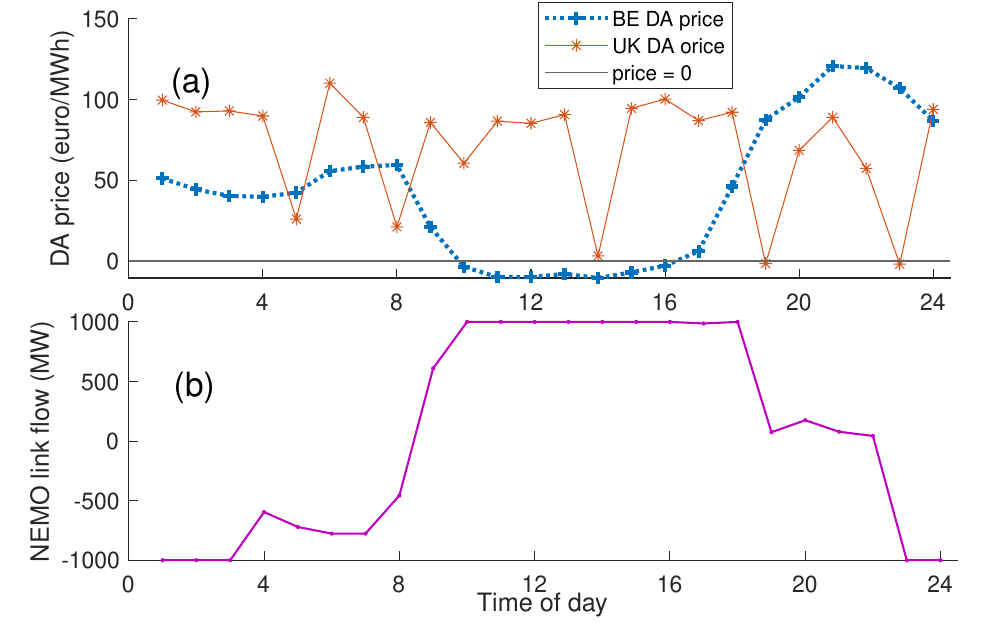}
	\vspace{-2pt}
	\caption{\small{(a) The DA prices for Belgium and the UK for 30th June 2024 and (b) stylized interconnector flow.}}
	\label{case1fig}
\end{figure}

In this case study, single and multi-region energy arbitrage models are evaluated for 30th June 2024. This day is selected due to prolonged negative price occurrence, see Figure \ref{case1fig}.

\begin{table}[!htbp]
\scriptsize
\centering
\caption{\small{Comparing arbitrage models for one and multiple regions for 1 day}}
\begin{tblr}{
  width = \linewidth,
  colspec = {Q[198]Q[133]Q[108]Q[144]Q[106]Q[117]Q[100]},
  cell{3}{1} = {r=8}{},
  cell{3}{2} = {r=4}{},
  cell{7}{2} = {r=4}{},
  cell{11}{1} = {r=3}{},
  cell{11}{2} = {r=3}{},
  cell{11}{3} = {r=3}{},
  cell{14}{1} = {r=3}{},
  cell{14}{2} = {r=3}{},
  cell{14}{3} = {r=3}{},
  vline{2} = {1-3,11,14}{},
  hline{3,11,14} = {-}{},
  hline{7} = {2-7}{},
}
Market        & Battery  & Model & NEMO      & Profit & Cycles & profit/ \\
participation & location &       & flow      & (euro) &        & cycle   \\
single        & BE       & LP    & n/a       & 136.1  & 2.43   & 56.0    \\
              &          & \textit{noDis} & n/a       & 134.2  & 1.96   & 68.3    \\
              &          & DP    & n/a       & 136.1  & 2.43   & 56.0    \\
              &          & MILP  & n/a       & 136.1  & 2.43   & 56.0    \\
              & UK       & LP    & n/a       & 236.5  & 2.77   & 85.4    \\
              &          & \textit{noDis} & n/a       & 236.5  & 2.77   & 85.4    \\
              &          & DP    & n/a       & 236.4  & 2.77   & 85.3    \\
              &          & MILP  & n/a       & 236.5  & 2.77   & 85.4    \\
dual          & BE       & MILP  & no flow   & 408.1  & 5.00   & 81.7    \\
              &          &       & with flow & 258.9  & 4.30   & 60.3    \\
              &          &       & reverse   & 393.4  & 4.60   & 85.5    \\
dual          & UK       & MILP  & no flow   & 425.5  & 5.00   & 85.2    \\
              &          &       & with flow & 331.9  & 4.49   & 74.0    \\
              &          &       & reverse   & 411.6  & 4.53   & 90.9    
\end{tblr}
\label{case1table}
\end{table}

Table \ref{case1table} benchmarks single and dual market EA models. The following are the key observations made:

$\bullet$ LP model for EA for single market proposed in \cite{hashmi2019optimal} provides solutions matching DP \cite{cheng2016co} and MILP \cite{mercier2023value} models. Although optimality under frequent occurrences of negative prices needs to be further evaluated.

$\bullet$ \textit{noDis} model's sub-optimality is evident for the battery located at the Belgian end. Note the \textit{noDis} model for UK matches other EA formulations as there are no occurrences of negative prices in the UK for the selected day (see Figure \ref{case1fig}).

$\bullet$ The MREA model generates up to two times more arbitrage revenue, clearly indicating the benefit of performing MREA. Further, revenue per cycle substantially improves for MREA, showing better battery utilization.

$\bullet$ For MREA, in this case, it is more beneficial for the battery to be located at the UK end compared to Belgium. 

$\bullet$ Interconnector flow has a substantial impact on the MREA gains. If the price potential and interconnector flow are in conflict, then EA gains are reduced. In the worst case, a 36\% reduction is observed.

\subsection{Case study 2: Profit assessment}
In this case study, long-term simulations are performed for day-ahead (DA) electricity prices in the UK and Belgium. 

Table \ref{case2table} presents the metrics for energy storage located at the Belgian and UK ends of the NEMO link. The aggregate financial revenue for placing a battery in Belgium or the UK is not much different, although there are year-to-year discrepancies. 

Note that $M_{\text{ind}}$ for all simulations is of the order of \(10^{-13}\), which is well below the typical numerical tolerance of the solver. This indicates that all simulations satisfy the criteria.
{Note that the minimum value for $x_{i}^{\text{ \sc a}} \cdot x_{i}^{\text{ \sc b}}$ is 0 based on the $P_{\text{MILP}}^{\text{MR}}$.}

\begin{figure}[!htbp]
	\center
	\includegraphics[width=2.75in]{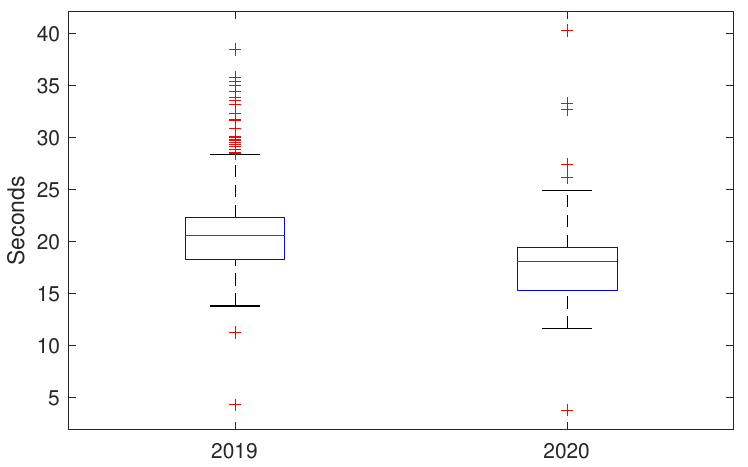}
	\vspace{-2pt}
	\caption{\small{Run time for 1000 MC simulations for 2019 and 2020.}}
	\label{fig:computationTime}
\end{figure}

Table \ref{case2table} also shows that the computation time for rolling-horizon simulations over 1 year is less than 30 seconds. This computational efficiency makes it suitable for real time implementation while considering stochasticity. Consolidating this claim, 1000 Monte Carlo (MC) simulations {are performed} for the years 2019 and 2020 {(with 1 day rolling horizon)}, see Figure \ref{fig:computationTime}. The mean runtime for these MC simulations is less than 25 seconds for the whole year.

\begin{table}
\scriptsize
\centering
\caption{\small{MREA for 8 years with battery located at BE or the UK for NEMO link}}
\begin{tblr}{
  width = \linewidth,
  colspec = {Q[129]Q[98]Q[150]Q[115]Q[154]Q[73]Q[190]},
  cell{2}{1} = {r=8}{},
  cell{12}{1} = {r=8}{},
  vline{2-3} = {1-2,10,12,20}{},
  vline{3} = {3-9,13-19}{},
  hline{2,10,12} = {-}{},
}
Location & Year  & Profit    & Cycles  & euro/cycle & time & $M_{\text{ind}}$           \\
BE       & 2024  & 40,034.4  & 890.3   & 45.0       & 39.3 & 3.1E-13 \\
         & 2023  & 33,654.0  & 726.5   & 46.3       & 25.9 & 8.5E-14 \\
         & 2022  & 95,545.3  & 979.4   & 97.6       & 67.1 & 9.2E-13  \\
         & 2021  & 52,004.5  & 805.6   & 64.6       & 29.1 & 1.9E-13 \\
         & 2020  & 12,464.7  & 698.4   & 17.8       & 17.5 & 1.8E-14  \\
         & 2019  & 11,374.7  & 637.9   & 17.8       & 13.9 & 1.5E-13 \\
         & 2018  & 16,564.5  & 743.2   & 22.3       & 20.0 & 6.9E-14  \\
         & 2017  & 16,753.8  & 800.0   & 20.9       & 22.0 & 2.3E-13 \\
         & Total & 278,395.8 & 6,281.4 & 41.5       & 29.3 & 0.0         \\
         &       &           &         &            &      &             \\
UK       & 2024  & 38,200.8  & 894.4   & 42.7       & 36.5 & 3.8E-13  \\
         & 2023  & 32,191.8  & 725.9   & 44.3       & 22.2 & 2.5E-14 \\
         & 2022  & 93,465.3  & 988.5   & 94.6       & 41.3 & 1.0E-13 \\
         & 2021  & 52,563.5  & 805.7   & 65.2       & 22.6 & 1.2E-14 \\
         & 2020  & 12,936.8  & 685.7   & 18.9       & 15.7 & 4.2E-14 \\
         & 2019  & 11,726.9  & 640.1   & 18.3       & 11.8 & 2.1E-11  \\
         & 2018  & 16,512.1  & 775.1   & 21.3       & 17.9 & 4.9E-14  \\
         & 2017  & 17,151.2  & 819.3   & 20.9       & 17.5 & 1.9E-13  \\
         & Total & 274,748.4 & 6,334.8 & 40.8       & 23.2 & 0.0         
\end{tblr}
\label{case2table}
\end{table}

\begin{table}
\scriptsize
\centering
\caption{\small{Multi and single EA models for 8 years of simulation}}
\begin{tblr}{
  width = \linewidth,
  colspec = {Q[104]Q[160]Q[165]Q[160]Q[160]Q[160]},
  cell{1}{2} = {c=2}{0.325\linewidth},
  cell{1}{4} = {c=3}{0.48\linewidth},
  vline{2-3} = {1}{},
  vline{2-6} = {2-12}{},
  hline{2-3,11-13} = {-}{},
}
      & MREA      &            & One market participation &           &           \\
YEAR  & Nominal   & with noDis & LP                       & LP noDis  & MILP      \\
2024  & 40,034.4  & 40,022.4   & 27,442.6                 & 27,439.5  & 27,456.0  \\
2023  & 33,654.0  & 33,624.0   & 25,686.5                 & 25,655.6  & 25,689.4  \\
2022  & 95,545.3  & 95,486.1   & 53,158.4                 & 53,096.1  & 53,169.3  \\
2021  & 52,004.5  & 51,978.7   & 22,383.5                 & 22,311.9  & 22,415.5  \\
2020  & 12,464.7  & 12,360.0   & 8,353.7                  & 8,320.5   & 8,382.4   \\
2019  & 11,374.7  & 11,349.6   & 8,001.6                  & 7,502.8   & 8,010.9   \\
2018  & 16,564.5  & 16,558.2   & 10,787.9                 & 10,783.4  & 10,787.9  \\
2017  & 16,753.8  & 16,735.1   & 9,045.4                  & 9,045.4   & 9,045.4   \\
Total & 278,395.8 & 278,114.1  & 164,859.6                & 164,155.3 & 164,956.7 \\
      & 100.0     & 99.9       & 59.2                     & 59.0      & 59.3      
\end{tblr}
\label{case2table2}
\end{table}

\begin{figure}[!htbp]
	\center
	\includegraphics[width=3.45in]{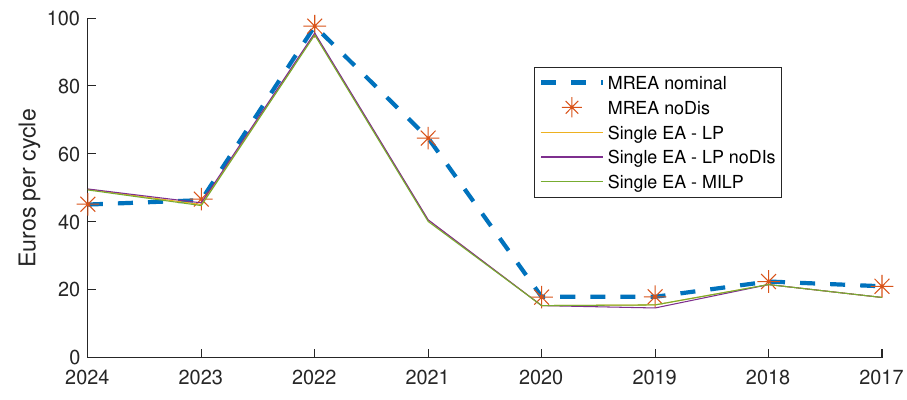}
	\caption{\small{Profit per 100\% cycle of battery operation.}}
	\label{case2fig2}
\end{figure}

Table \ref{case2table2} compares five models for total EA gains for a battery located at the Belgian end of the NEMO link interconnector. These 5 models are: (i) nominal MREA (denoted as $P_{\text{MILP}}^{\text{MR}}$ in this work), (ii) MREA with \textit{noDis}, (iii) LP, (iv) LP with \textit{noDis}, and (v) MILP models.
Models (iii)-(v) consider only the local Belgian market participation, while (i) and (ii) consider both the Belgian and the UK DA markets.
MREA outperforms the single market participation by more than 40\% additional revenue generation. 
However, from Figure \ref{case2fig2}, MREA does not substantially improve revenue per equivalent 100\% depth-of-discharge cycles of battery operation. Thus, there is a need for better utilization of such expensive grid-scale batteries.

\begin{mdframed}

{
Note that considering the per kWh cost of 100 euros, the total battery cost of 1 MWh (0.5C-0.5C)is 100,000 euros, which has 178\% return for $P_{\text{MILP}}^{\text{MR}}$ model without reaching the storage cycle life of 7200 cycles. Refer to the battery parameters in Table \ref{tab:batparameters}.
}
    
\end{mdframed}

\subsection{Case study 3}
This case study aims to improve the utilization of grid-scale batteries. Batteries have a limited cycle life, necessitating their better use. Previously,  $\eta_{\text{pseudo}}$ {is introduced} as a mechanism to limit less profitable cycles of battery, see \cite{hashmi2018limiting} for a detailed explanation.

\begin{figure}[!htbp]
	\center
	\includegraphics[width=3.4in]{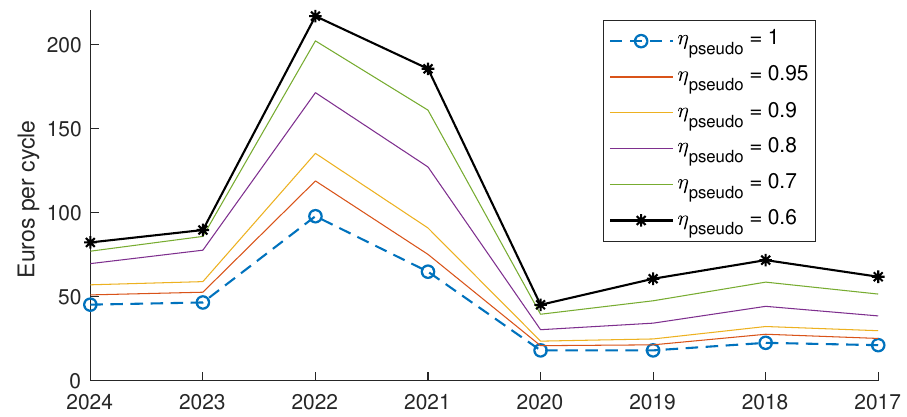}
	\vspace{2pt}
	\caption{\small{Limiting cycles of operation.}}
	\label{case3fig}
\end{figure}

\begin{figure}[!htbp]
	\center
	\includegraphics[width=3.3in]{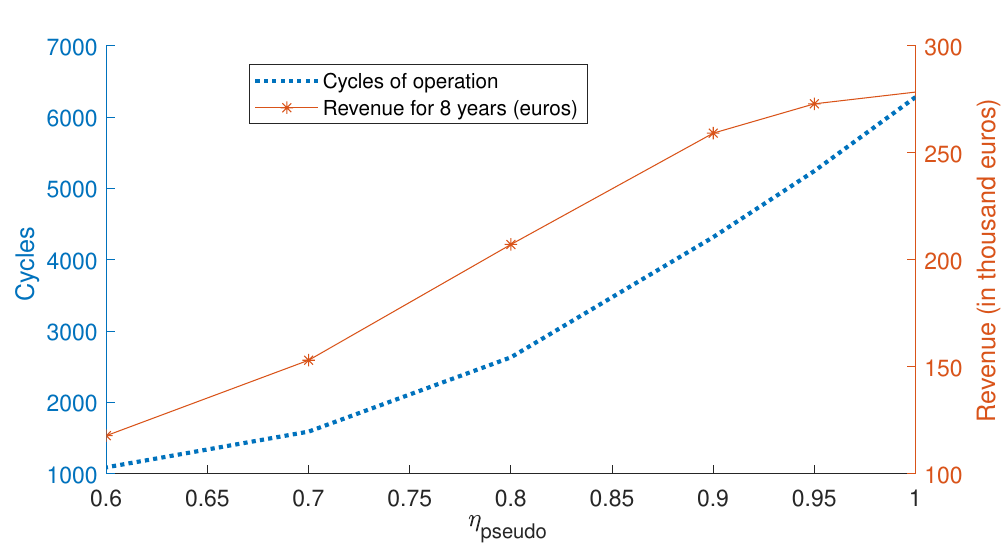}
	\vspace{-2pt}
	\caption{\small{Comparing cycles of operation with revenue for different $\eta_{\text{pseudo}}$.}}
	\label{case3fig2}
\end{figure}

Figure \ref{case3fig} shows the substantial improvement in revenue per cycle with a reduction in $\eta_{\text{pseudo}}$. For $\eta_{\text{pseudo}} = 0.7$, the revenue reduced from 278k euros to 153k euros ($\approx 45\%$ reduction), while the cycles of operation reduced from 6281 to 1593 ($\approx 74\%$ reduction). This is also shown in Figure \ref{case3fig2}.

\subsection*{Online repository}
{
The MILP-based MREA,  $P_{\text{MILP}}^{\text{MR}}$,
described in this paper, is publicly available at
\href{https://github.com/umar-hashmi/Inter-Refional-Energy-Arbitrage}{{github.com/umar-hashmi/Inter-Refional-Energy-Arbitrage}}. 
}

\section{Conclusion and future work}
\label{section5}
This paper successfully developed and validated a novel Mixed-Integer Linear Programming (MILP) model for Multi-Region Energy Arbitrage (MREA), enabling a single battery energy storage system to co-optimize its operation across two interconnected electricity markets. By employing a disjunctive linearization technique, {this work} exactly reformulated the non-convex constraint that prohibits simultaneous charging and discharging, resulting in a computationally efficient and tractable model.

The case study, based on 8 years of historical day-ahead market data for the NEMO Link interconnector between Belgium and the UK, demonstrates the significant financial benefits of MREA. Our results show that participating in both markets can increase arbitrage revenue by over 40\% compared to operating in a single local market. The analysis also highlighted the critical impact of interconnector congestion, which can reduce potential gains by up to 36\% if the flow opposes the optimal arbitrage strategy. Furthermore, a mechanism to manage battery degradation by limiting less profitable cycles {is introduced}, showcasing a practical trade-off that can improve revenue per cycle by sacrificing a portion of the total revenue to extend the asset's operational life. The model's computational efficiency, with annual simulations completing in under 30 seconds, confirms its suitability for practical applications, including near-real-time decision-making.

For future work, several extensions to this research can be explored. First, while the proposed formulation is extensible, its application and performance should be evaluated for scenarios involving more than two interconnected markets. Second, incorporating uncertainty is a crucial next step; the deterministic model could be advanced to a stochastic programming framework to account for unforeseen fluctuations in market prices, interconnector availability, and renewable energy generation. Finally, the scope of services could be expanded to include the co-optimization of energy arbitrage with ancillary services, such as frequency regulation, across multiple regions, which would provide a more holistic valuation of energy storage in interconnected power systems.

\section*{Acknowledgement}
This work is supported by the 
Flemish Government and Flanders Innovation \& Entrepreneurship (VLAIO) through 
IMPROcap project (HBC.2022.0733)
and by the Flemish Government within the project Innovative solutions for under-ground high-voltage lines and grids (Etch).

\bibliographystyle{IEEEtran}
\bibliography{reference}

\end{document}